\documentclass[%
reprint,
 amsmath,amssymb,
prb
]{revtex4-2}

\usepackage{graphicx}% Include figure files
\usepackage{dcolumn}% Align table columns on decimal point

\usepackage[labelfont=bf]{caption}
\usepackage{bm}% bold math
\usepackage[english]{babel}
\usepackage[utf8]{inputenc}
\DeclareUnicodeCharacter{2212}{-} % to avoid latex error from inputenc package
\DeclareUnicodeCharacter{03B4}{\ensuremath{\delta}} 
\usepackage{siunitx} % SI units
\usepackage{textcomp} % latex symbols
\usepackage{textcase} % provides \NoCaseChange for titles/headers that get uppercased
\DeclareSIUnit\angstrom{\text{\AA}} % Angstrom SI unit
\usepackage{enumitem} % make custom enumerator lists
\usepackage{threeparttable}

\usepackage[table,xcdraw,svgnames]{xcolor} % more colors
\usepackage[pdfencoding=auto,psdextra,colorlinks]{hyperref} % hyperreferences in document
\AtBeginDocument{\hypersetup{ % needs to be defined at start
  colorlinks   = true, %Colours links instead of ugly boxes
  urlcolor     = purple, %Colour for external hyperlinks
  linkcolor    = teal, %Colour of internal links
  citecolor   = violet %Colour of citations
}}

\usepackage{xcolor, soul}
\sethlcolor{yellow}

\usepackage{subcaption} % subfigures
\usepackage[most]{tcolorbox}
\usepackage{xcolor}

\usepackage{tabularx} %% table cell width automtically adjusted
\usepackage{booktabs} %% nicer tables (bottom / top line)
\usepackage{multirow} %% combine several cells
\usepackage{makecell}

\usepackage{duckuments}

\begin{document}

% Box setup
\tcbset{
  graybox/.style={
    colback=gray!20,   % Light gray background
    colframe=gray!20,  % Same as background → invisible border
    boxrule=0pt,       % No border thickness
    arc=1mm,           % Optional: slightly rounded corners
    left=2mm, right=2mm, top=1mm, bottom=1mm, % Optional padding
  }
}

\tcbset{
  orangebox/.style={
    colback=orange!20,   % Light gray background
    colframe=orange!20,  % Same as background → invisible border
    boxrule=0pt,       % No border thickness
    arc=1mm,           % Optional: slightly rounded corners
    left=2mm, right=2mm, top=1mm, bottom=1mm, % Optional padding
  }
}

\title{
Interfacial electrostatics: From polar insulators to charged-layer metals 
}

\author{Edith Simmen$^1$}
    \email{edith.simmen@mat.ethz.ch}
\author{Nicholas C. Bristowe$^2$}
\author{Philippe Ghosez$^3$}
\author{Nicola A. Spaldin$^1$}%
\affiliation{%
$^1$ Materials Theory, ETH Zurich, 8093 Zurich, Switzerland, 
$^2$ Condensed Matter Physics, Physics Department, Durham University, Durham, United Kingdom, 
$^3$ Theoretical Materials Physics, Q-MAT, Université de Liège, B-4000 Sart-Tilman, Belgium
}%

\date{\today}

\begin{abstract}
Polar discontinuities are typically described as arising from charged atomic planes, a spontaneous polarization, or a combination, in insulating films interfaced with vacuum, nonpolar insulators or metals. The resulting electrostatics are well established for systems such as LaAlO$_3$/SrTiO$_3$, BaTiO$_3$/SrRuO$_3$, and BiFeO$_3$/SrRuO$_3$. However, recent observations of ``polar-discontinuity-like'' behavior at interfaces in which the metal contains formally charged layers (e.g., rare-earth nickelates) challenge this picture. In this work, we summarize the theoretical framework used to define and quantify interface charge, layer polarization, and screening in the three reference systems above, and then discuss signatures attributed to the formal-charge polar discontinuity arising when SmNiO$_3$, a metal with formally charged (001) layers, is interfaced with ferroelectric BaTiO$_3$. We show how concepts developed for polar discontinuities between two insulators and for screening at metallic interfaces can rationalize the reported effects and identify where our understanding of these systems leaves open questions.
\end{abstract}

\maketitle

\section{Introduction}
Interfacial phenomena have repeatedly attracted considerable attention from the perovskite oxide community, both to develop next-generation electronic devices, and because they can lead to non-trivial and intriguing physics. Notably, uncompensated interface charges can cause a variety of effects at the interface or in the interfacing materials. Such charges lead to the formation of a two-dimensional electron gas (2DEG) in LaAlO$_3$/SrTiO$_3$ heterostructures \cite{bristowe_origin_2014, lemal_polarityfield_2020, ohtomo_highmobility_2004, nakagawa_why_2006}, the presence of a critical thickness for the spontaneous polarization in ferroelectric thin films \cite{junquera_critical_2003, mehta_depolarization_1973, efe_engineering_2024}, or unusual ferroelectric topological textures \cite{yadav_observation_2016,das_observation_2019,junquera_topological_2023}.

Interface charges can arise when there is a change in the value of the polarization perpendicular to the interface -- a so-called polar discontinuity -- and their origin and screening have been extensively studied in theory and experiments~\cite{ohtomo_highmobility_2004, nakagawa_why_2006, stengel_electrostatic_2011, junquera_critical_2003, black_electricfield_1999}. Within the modern theory of polarization, a polar discontinuity occurs at an interface between materials with different bulk polarizations~\cite{king-smith_theory_1993, resta_macroscopic_1993, vanderbilt_electric_1993a, stengel_electrostatic_2011}. For a (001) interface, the spontaneous polarization of ferroelectrics such as BaTiO$_3$ contributes to the bulk polarization and hence to the bound interface charge. In centrosymmetric materials such as LaAlO$_3$, formally charged atomic (001) planes can likewise produce a polar discontinuity at the interface with vacuum or a material with uncharged layers. The screening of these interface charges can be modeled via an effective screening length at metal interfaces~\cite{ghosez_firstprinciples_2006, black_electricfield_1999, mehta_depolarization_1973} or by accounting for the energetic cost of generating screening charges via other compensation mechanisms \cite{bristowe_origin_2014, hong_screening_2016}.  

More recently, effects consistent with a polar discontinuity have been observed and associated with metals with formally charged layers, particularly for metallic rare-earth nickelates \cite{tao_ferroelectricity_2016, wu_interface_2014, malashevich_controlling_2018, simmen_interplay_2025, stylianidis_alternative_, hwang_polarizationcontrolled_2025, geisler_design_2017, efe_polarizing_2026, yang_ferroelectric_2019, marshall_conduction_2014} or La$_{1-x}$Sr$_\text{x}$MnO$_{3}$ \cite{yu_interface_2012, chang_atomically_2011, wang_atomicscale_2019, li_enhancement_2022, strkalj_insitu_2020, burton_evolution_2010, mundy_visualizing_2014, peters_polarization_2020} interfaced with BaTiO$_3$ or SrTiO$_3$. While the behavior is consistent with a formal-charge polar discontinuity, it is unclear how formally charged layers can persist in a metal with free carriers \cite{yu_interface_2012, chang_atomically_2011, wang_atomicscale_2019, li_enhancement_2022, strkalj_insitu_2020}, given our understanding of basic electrostatics. 

In this perspective article, we summarize the current state of research of heterostructures with polar discontinuities and highlight the fundamental questions that remain unanswered within the current theoretical framework. Following four representative systems (see summary in Tab. \ref{tab:summary_properties}), we will first review the concepts governing the current understanding of polar discontinuities between insulators, ferroelectrics and metallic electrodes before rationalizing the apparent formal-charge polar discontinuity in systems where the formally charged layers reside in the metal.
\begin{itemize}
    \item[--] LaAlO$_3$/SrTiO$_3$: The interface between these two paraelectric insulators has been widely studied due to the 2DEG that forms to screen the polar discontinuity \cite{ohtomo_highmobility_2004, goniakowski_polarity_2008, bristowe_origin_2014}.
    \item[--] BaTiO$_3$/SrRuO$_3$: The interface of the prototypical ferroelectric BaTiO$_3$ to the metal SrRuO$_3$ has been essential in understanding incomplete screening by the electrode in perovskite oxide systems \cite{junquera_critical_2003, gerra_ionic_2006, stengel_enhancement_2009}.
    \item[--] BiFeO$_3$/SrRuO$_3$: The ferroelectric and III-III perovskite BiFeO$_3$ is one of the few studied systems where a spontaneous polarization and charged layers coexist within the same material \cite{efe_happiness_2021, spaldin_layer_2021, efe_engineering_2024, begon-lours_factors_2018}. 
    \item[--] SmNiO$_3$/BaTiO$_3$: This system combines SmNiO$_3$, a representative of the rare-earth nickelate family that has formally charged layers in the metallic phase, with the prototypical ferroelectric BaTiO$_3$.  
\end{itemize}

Our focus is on the theoretical concepts and first-principles modeling of systems with a formal-charge polar discontinuity. More detailed reviews can be found in Ref. \cite{efe_engineering_2024, vaz_epitaxial_2021} for related experimental findings, Ref. \cite{goniakowski_polarity_2008, stengel_electrostatic_2011, hong_screening_2016} for polar discontinuities between insulators and Ref. \cite{ghosez_firstprinciples_2006, stengel_band_2011, vaz_epitaxial_2021} for an extensive review of metal and ferroelectric interfaces.

\begin{table}[h]
    \centering
    \caption{Properties of the four representative systems discussed here. The abbreviations PE, FE and M refer to materials that are paraelectric, ferroelectric or metals. We calculate the (001) layer charges based on the formal charges. }
    \label{tab:summary_properties}
    \renewcommand{\arraystretch}{1.2}
    \begin{tabular}{r@{/}l @{$\hspace{1em}$} r@{/}l @{$\hspace{1em}$} r@{/}l @{$\hspace{1em}$} c@{/}l}
        \hline
        \multicolumn{2}{c}{\textbf{Materials}} &
        \multicolumn{2}{c}{\thead{Layer \\ charges}} &
        \multicolumn{2}{c}{\thead{Polar \\ distortions}} &
        \multicolumn{2}{c}{\thead{Interface \\ type}} \\
        \hline
        LaAlO$_3$ & SrTiO$_3$ & $\pm1e$ & 0 & 0 & 0 & PE & PE \\
        BaTiO$_3$ & SrRuO$_3$ & 0 & 0 & $\pm P_\text{spont}$ & 0 & FE & M \\
        BiFeO$_3$ & SrRuO$_3$ & $\pm1e$ & 0 & $\pm P_\text{spont}$ & 0 & FE & M \\
        SmNiO$_3$ & BaTiO$_3$ & $\pm1e$ & 0 & 0 & $\pm P_\text{spont}$ & M & FE \\
        \hline
    \end{tabular}
\end{table}

\section{Charged layers and/or spontaneous polarization in the insulating slab}

\begin{figure*}[t]
    \centering
    \includegraphics[width=\textwidth]{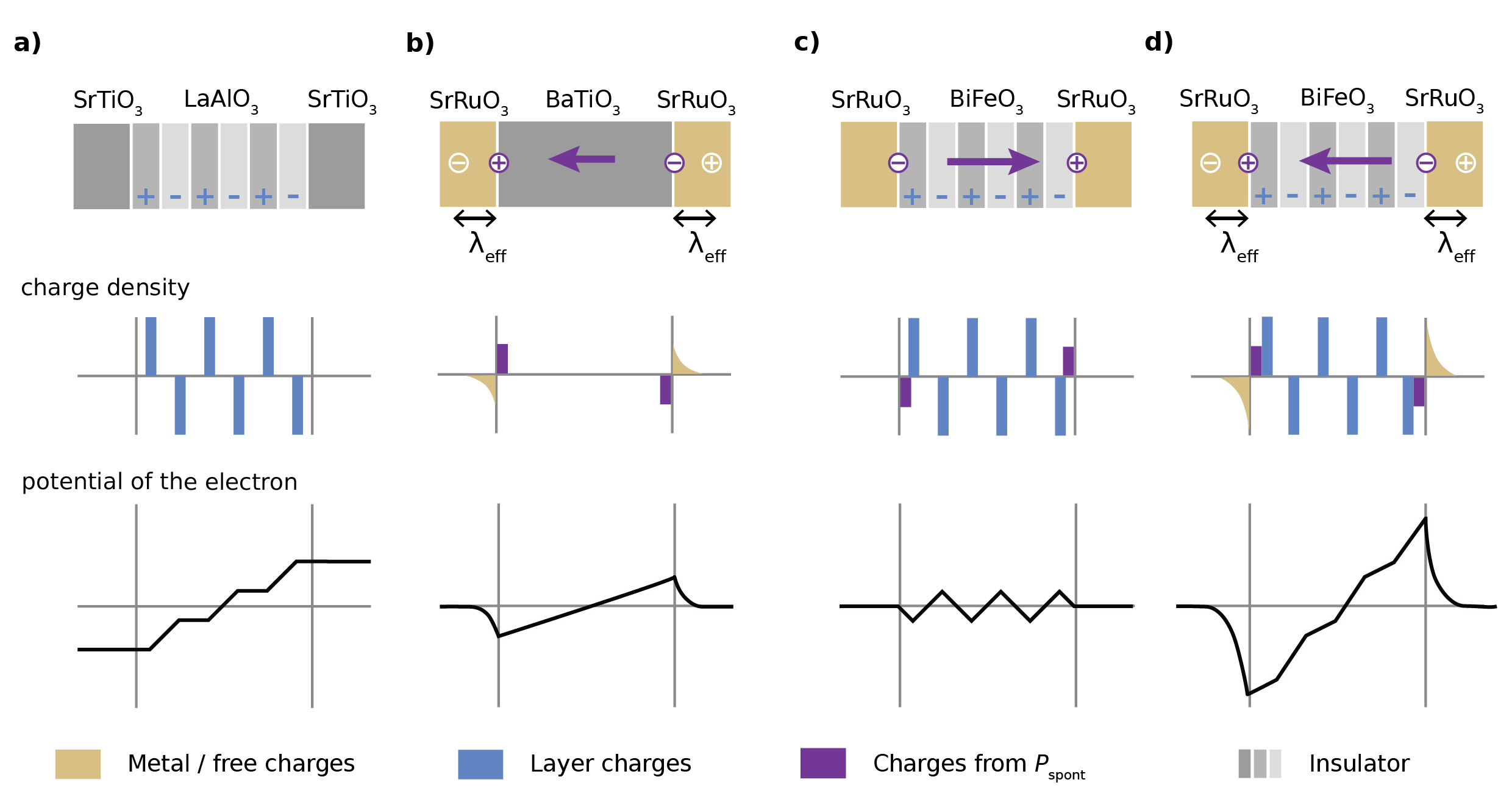}
    \caption{Illustration of the interface, charge density and electrostatic potential for a) asymmetric SrTiO$_3$/LaAlO$_3$, b) symmetric SrRuO$_3$/BaTiO$_3$ and c, d) asymmetric SrRuO$_3$/BiFeO$_3$. The metallic part of the structure and free carriers are shown in yellow, insulating materials in gray, and bound charges from a layer polarization and spontaneous polarization in blue or purple, respectively. We assume short-circuit boundary conditions for the capacitor structures (b-d).}
    \label{fig:illu_all}
\end{figure*}

\subsection{LaAlO$_3$/SrTiO$_3$ interface}

In this section, we introduce the concepts that we will build on throughout this work by analyzing the polar discontinuity arising between two insulators, LaAlO$_3$ and SrTiO$_3$, that have differently charged layers. The resulting 2DEG at this interface has been thoroughly studied, both theoretically and experimentally \cite{ohtomo_highmobility_2004, nakagawa_why_2006, bristowe_origin_2014, lemal_polarityfield_2020}, owing to its superconducting \cite{reyren_superconducting_2007, bert_direct_2011, li_coexistence_2011}, magnetic \cite{brinkman_magnetic_2007, bert_direct_2011, li_coexistence_2011}, and other interesting properties \cite{caviglia_tunable_2010, caviglia_electric_2008}. We refer the reader to the literature for an extensive summary of LaAlO$_3$/SrTiO$_3$  \cite{bristowe_origin_2014}, and for an in-depth discussion of the theory behind polar discontinuities \cite{stengel_electrostatic_2011}. 

\subsubsection{Layer polarization and its relation to interface charge}
\label{sec:player_lao}

In an ionic solid, layer charges are determined by dividing the unit cell into sublayers (AO and BO$_2$ for an ABO$_3$ perovskite along [001]) and adding up the formal charges on the ions of these sublayers \cite{tasker_stability_1979, stengel_electrostatic_2011}. Since LaAlO$_3$ is a III-III perovskite with  formal charges La$^{3+}$, Al$^{3+}$ and O$^{2-}$, it has charged layers of (LaO)$^{+}$ and (AlO$_2)^{-}$ along the [001] direction. The II-IV perovskite SrTiO$_3$ consists of Sr$^{2+}$, Ti$^{4+}$ and O$^{2-}$ ions and has neutral sublayers along [001]. The polarization formed by these charged layers is referred to as `layer polarization' \cite{spaldin_layer_2021, efe_happiness_2021, gattinoni_prediction_2022} or `formal polarization' \cite{stengel_berryphase_2009, bristowe_origin_2014, fang_how_2023} in the literature; here, we will use the term `layer polarization' ($\mathbf{P}_\text{layer}$). Note that, although the layer polarization of a II-IV perovskite is zero along [001], the sublayers can be charged along another direction (e.g. [111]).

The concept of layer polarization maps formally onto the modern theory of polarization \cite{king-smith_theory_1993, resta_macroscopic_1993, vanderbilt_electric_1993a} via the Wannier functions in an insulator \cite{stengel_electrostatic_2011}. The total bulk polarization is defined modulo a polarization quantum via the following formula in terms of Wannier functions
\begin{equation}
    \mathbf{P}_\text{bulk} = \frac{1}{\Omega} \left( \sum_{i=1}^{N_\text{ions}} \mathbf{R}_i Z_i - 2 e \sum_{j=1}^{N_e/2} \mathbf{r}_j \right),
    \label{eq:P_layer}
\end{equation}
where $\mathbf{R}_i$ and $Z_i$ are the position and the valence charge of ion $i$, respectively, $\mathbf{r}_j$ is the position of Wannier function $j$ and $e$ the elementary charge. The bulk polarization is defined modulo a polarization quantum $\mathbf{P}_\text{q} = \frac{e \mathbf{R}}{\Omega}$, where $e$ is the elementary charge and $\mathbf{R}$ are the lattice vectors of the unit cell with volume $\Omega$. The bulk polarization thus corresponds to a polarization lattice with allowed values spaced by $\textbf{P}_\text{q}$. This results in a zero-containing polarization lattice with the values $P^\text{[001]}_\text{SrTiO$_3$} = \SI{0}{} \pm n \cdot P^\text{[001]}_\text{q, SrTiO$_3$}$ for SrTiO$_3$ and a half-quantum containing lattice with values $P^\text{[001]}_\text{LaAlO$_3$} = \SI{0.5}{\elementarycharge\per\textit{S}} \pm n \cdot P^\text{[001]}_\text{q, LaAlO$_3$}$ \cite{stengel_electrostatic_2011} for LaAlO$_3$, where $S$ is the area of the (001) plane of the unit cell. For centrosymmetric structures, we can use Eq.~\ref{eq:P_layer} to obtain the layer polarization ($\mathbf{P}_\text{bulk} = \mathbf{P}_\text{layer}$). Note that the same result for $\mathbf{P}_\text{bulk}$ is obtained using a Berry phase calculation \cite{stengel_berryphase_2009, stengel_electrostatic_2011}.  

Using Wannier functions, we can also define the layer charge introduced at the beginning of this section more rigorously. If a solid has some degree of ionicity, the Wannier functions can be assigned to their closest ion core and summed per AO and BO$_2$ sublayer, resulting in the layer charge. By assigning the Wannier functions to the closest ion, we obtain Wannier ions with charge $Q_i$ at the position of the original ion core, resulting in a physically more meaningful separation between electrons and ions. In comparison, the distinction between ions and electrons introduced in Eq.~\ref{eq:P_layer} is arbitrary and not physically meaningful \cite{stengel_electrostatic_2011}. Finally, we can reformulate the bulk polarization in terms of these Wannier ions \cite{stengel_electrostatic_2011}.
\begin{equation}
    \mathbf{P}_\text{bulk} = \frac{1}{\Omega} \sum_{i=1}^{N_\text{ions}} \left( \mathbf{R}_i Q_i - 2 e \sum_{j \in i} \left( \mathbf{r}_j - \mathbf{R}_i \right) \right)
    \label{eq:P_WI}
\end{equation}
Here, the inner sum corresponds to the difference in positions between the Wannier center $\mathbf{r}_j$ and its closest ionic core $\mathbf{R}_i$. The sum over these dipole moments accounts for the contribution of the electron-cloud deformation to the polarization. Since the sum over the dipole moment of all ions is zero in centrosymmetric systems, Eq.~\ref{eq:P_WI} corresponds to a summation over ions with formal charges $Q_i$, explaining why we usually obtain the correct results for the layer polarization using formal charges \cite{stengel_electrostatic_2011}. 

As long as the Wannier functions can still be unambiguously assigned to the same ion core, the value of the layer charge does not change with covalency. If the covalency in a material increases, the Wannier center moves towards the bond center but, nevertheless, remains closest to the same ion core \cite{stengel_electrostatic_2011}. More covalency in this case only results in the charge being distributed differently within the bonding volume which, according to Gauss's law, does not affect the amount of enclosed charge \cite{bristowe_net_2011}. This absent influence of covalency on the layer charge has been extensively discussed for LaAlO$_3$/SrTiO$_3$ \cite{bristowe_net_2011} or more generally in Ref.~\cite{stengel_electrostatic_2011} (for a discussion of a fully covalent system, also see Ref. \cite{stengel_electrostatic_2011}).

The bulk polarizations of LaAlO$_3$ ($\mathbf{P}_\text{bulk,LAO}$) and SrTiO$_3$ ($\mathbf{P}_\text{bulk,STO}$) calculated with Eq.~\ref{eq:P_layer} are directly linked to the interface charge $\sigma_\text{int}$ via the `interface theorem' \cite{vanderbilt_electric_1993a} 
\begin{equation}
    \sigma_\text{int} = \left( \mathbf{P}_\text{bulk,LAO} - \mathbf{P}_\text{bulk,STO} \right) \cdot \mathbf{n},
    \label{eq:interface_theorem} 
\end{equation}
where $\mathbf{n}$ is the surface normal. If both materials have the same bulk polarization, the interface charge is zero. However, since SrTiO$_3$ and LaAlO$_3$ have different polarization lattices and thus a different layer polarization, flat (001) surfaces have an interface charge of $\SI{\pm0.5}{\elementarycharge\per\textit{S}}$ \cite{bristowe_net_2011}. Note that while I-V and III-III perovskites both have a half-quantum containing polarization lattice, a flat (001) interface between them has a bound charge of $\SI{\pm1}{\elementarycharge\per\textit{S}}$ since their AO and BO$_2$ sublayers are oppositely charged. 

\subsubsection{Formation of a 2DEG}

The polar discontinuity resulting from the polarization lattice mismatch between LaAlO$_3$ and SrTiO$_3$ leads to an unfavorable potential divergence inside LaAlO$_3$, SrTiO$_3$ or both. The oppositely charged layers of LaAlO$_3$ (blue +/–) and the resulting charge density (blue bars) are illustrated in Fig.~\ref{fig:illu_all}a for stoichiometric LaAlO$_3$ sandwiched between SrTiO$_3$ under open-circuit boundary conditions. The charged (LaO)$^+$ and (AlO$_2$)$^-$ sublayers at the respective LaO-TiO$_2$ and AlO$_2$-SrO interfaces are not compensated, resulting in a nonzero electric field and a respective increase or decrease in potential between each pair of sublayers, see bottom of Fig.~\ref{fig:illu_all}a. Unless compensated, this potential increases linearly throughout the thickness of the film and diverges. 

The polar discontinuities at the LaO-TiO$_2$ and the AlO$_2$-SrO interface can be fully screened by \SI[explicit-sign=+]{0.5}{\elementarycharge\per\textit{S}} and \SI{-0.5}{\elementarycharge\per\textit{S}} screening charges, respectively. In LaAlO$_3$/SrTiO$_3$, these screening charges are provided by a 2DEG and/or ionic defects at the interface  \cite{ohtomo_highmobility_2004}. They can be generated through different mechanisms such as charge transfer between the interfaces, i.e. Zener tunneling (intrinsic screening), or defect formation (extrinsic mechanism) \cite{goniakowski_polarity_2008, hong_screening_2016, nakagawa_why_2006, yu_polarityinduced_2014, bristowe_origin_2014, kalabukhov_improved_2011}, where oxygen vacancies are found to be the main source of defects that provide free electrons for the 2DEG \cite{bristowe_origin_2014, bristowe_surface_2011, lemal_polarityfield_2020, herranz_high_2007a}. 

\begin{figure*}[t]
    \centering
    \includegraphics[width=\textwidth]{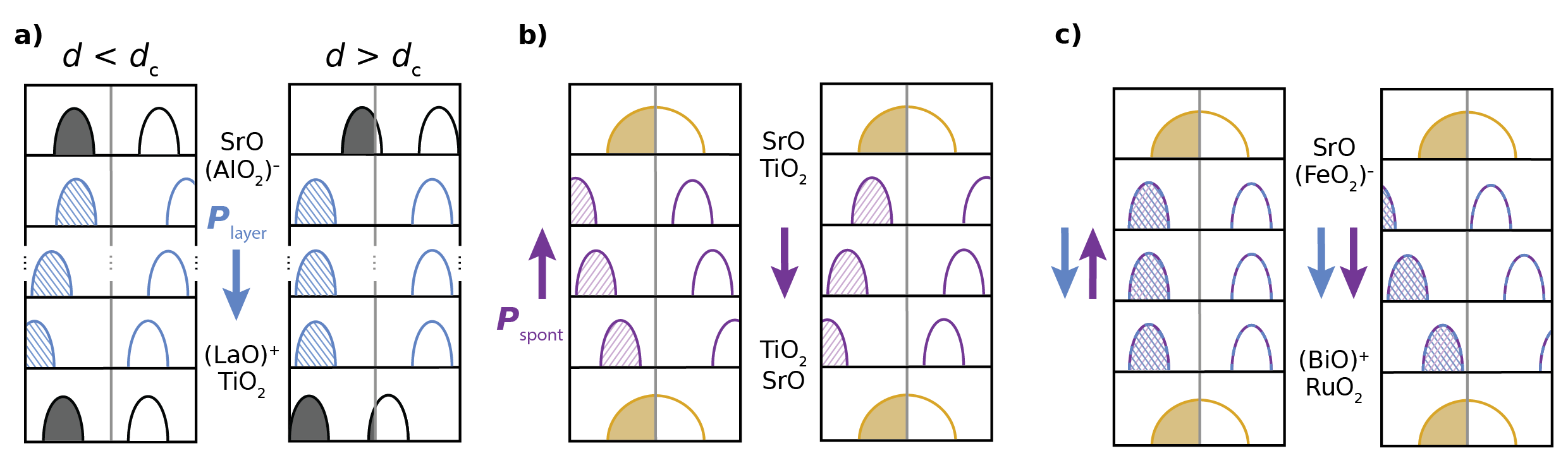}
    \caption{Illustration of the layer-projected DOS for a) asymmetric LaAlO$_3$/SrTiO$_3$ below (left) and above (right) the critical thickness, and the up (left) or down (right) direction of the spontaneous polarization in b) symmetric BaTiO$_3$/SrRuO$_3$ and c) asymmetric BiFeO$_3$/SrRuO$_3$. Bands with shifts due to the layer and spontaneous polarization are shown in blue and purple; bands from SrTiO$_3$ and SrRuO$_3$ are shown in gray and yellow, respectively.}
    \label{fig:illu_dos}
\end{figure*}

Lastly, the polar discontinuity and screening by the 2DEG lead to specific features in the density-of-states (DOS), see illustration in Fig.~\ref{fig:illu_dos}a. Charge transfer between the interfaces can occur in films thicker than the critical thickness $d_\text{c}$ \cite{bristowe_oxide_2009}. The potential difference between the two interfaces grows with the thickness of the LaAlO$_3$ slab (left in Fig.~\ref{fig:illu_dos}a) and charge transfer occurs once this difference is larger than the effective band gap of the system (this corresponds to the band gap of SrTiO$_3$ for type-I band alignment). Above the critical thickness, the valence band crosses the Fermi level to generate holes at the negative interface and the conduction band moves below the Fermi level to generate free electrons at the positive interface \cite{bristowe_oxide_2009}, see right side of Fig.~\ref{fig:illu_dos}a.  The potential difference is visible in the DOS, leading to very different band alignments for the two interfaces \cite{bristowe_oxide_2009}. Note that the illustration shows complete screening of the interface charge with zero field across the LaAlO$_3$ slab. In reality, a finite field is usually present since the amount of charge transferred depends on the slab thickness and the interfaces are chemically different.

In conclusion, if two centrosymmetric materials have differently charged layers in the bulk, they form a polar discontinuity at the interface. This polar discontinuity can be screened by the formation of a 2DEG, by transferring charge between the interfaces and/or by forming charged defects. Apart from LaAlO$_3$/SrTiO$_3$, polar discontinuities have also been studied at the interface of various other insulating perovskites: LiInO$_3$/BaSnO$_3$ \cite{fang_how_2023, aggoune_tuning_2021}, LaMnO$_3$/SrTiO$_3$ \cite{miyagawa_nondestructive_2025} \cite{chen_electron_2017} and GdTiO$_3$/SrTiO$_3$ \cite{banerjee_electronic_2015} are examples of other III-III/II-IV interfaces while LaAlO$_3$/A$^+$B$^{5+}$O$_3$ \cite{wang_creating_2016} is an example for a III-III/I-V and LaTiO$_3$/KTaO$_3$ \cite{zou_latio3_2015} and KNbO$_3$/SrTiO$_3$ \cite{murray_theoretical_2009} for a  I-V/II-IV system.

\subsection{Interface between BaTiO$_3$ and SrRuO$_3$}
\label{sec:bto_sro}

\begin{figure}
    \centering
    \includegraphics[width=\linewidth]{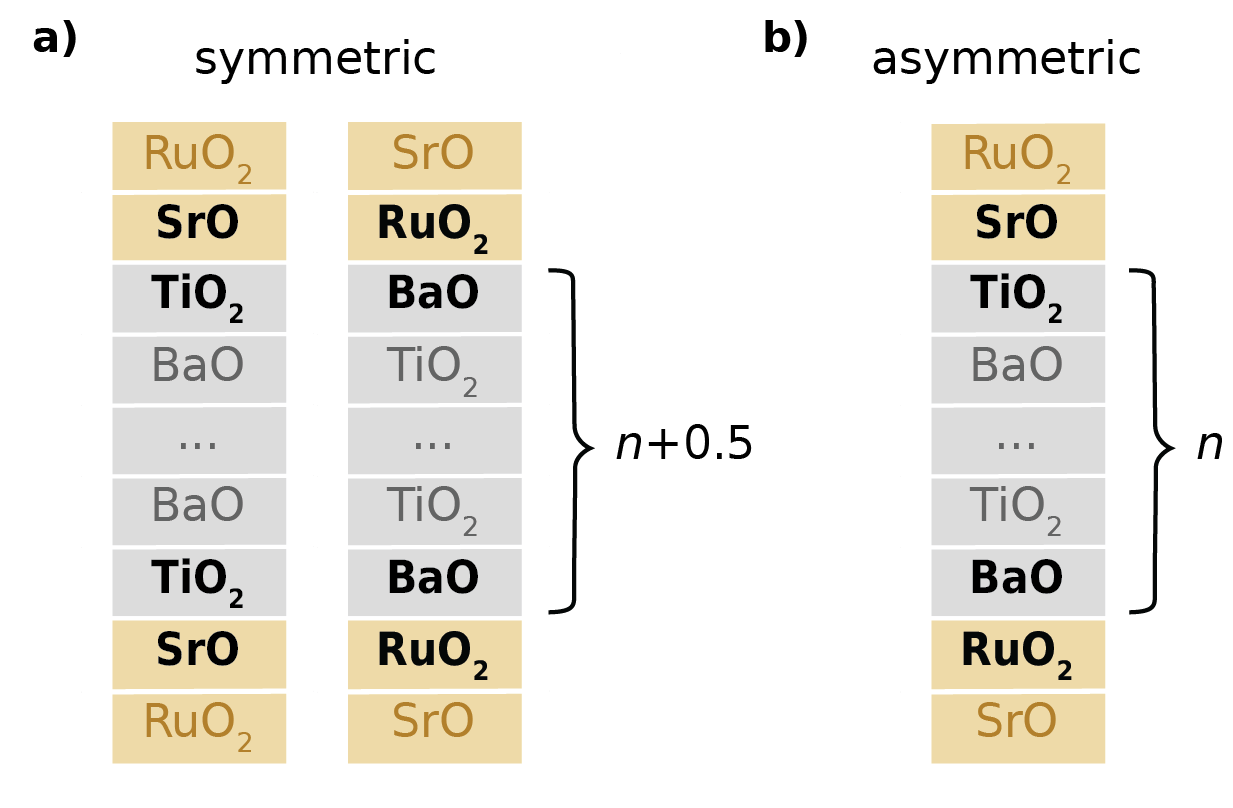}
    \caption{Capacitor with a) symmetric or b) asymmetric interfaces. For symmetric capacitors, the BaTiO$_3$ and SmNiO$_3$ slabs are non-stoichiometric ($n+0.5$ BaTiO$_3$ formula units) and two different interface choices are possible.}
    \label{fig:symm_as_sl}
\end{figure}

Polar discontinuities also occur at the surface or interface of ferroelectrics with non-polar materials, e.g. between ferroelectric BaTiO$_3$ and centrosymmetric SrTiO$_3$. In this case, both materials are formally II-IV perovskites and the polar discontinuity originates only from a difference in spontaneous polarization. Since ferroelectrics are often combined with metallic electrodes, we will use the prototypical ferroelectric BaTiO$_3$ with the metallic II-IV SrRuO$_3$ as our model system. In this section, we will first highlight some of the differences between spontaneous and layer polarization and then discuss the role of metallic screening by the electrode. We initially consider only symmetric BaTiO$_3$/SrRuO$_3$ capacitors, see illustration in Fig.~\ref{fig:symm_as_sl}a, and discuss the effect of asymmetric interfaces in the end (Fig.~\ref{fig:symm_as_sl}b).

The spontaneous polarization, as opposed to the layer polarization, originates from polar distortions in the material and can take values other than an integer multiple of the polarization quantum (modulo the polarization quantum). Its magnitude depends on the polar displacements $\Delta \mathbf{r}_i$ of the ions away from their centrosymmetric position and the corresponding deformation of the electronic clouds. The spontaneous polarization can only be measured as a change in polarization with respect to a non-polar reference, with the bulk polarization of the polar ($ \mathbf{P}_\text{bulk,pol}$) and non-polar ($\mathbf{P}_\text{bulk,npol}$) structures defined as in Eq.~\ref{eq:P_layer} \cite{king-smith_theory_1993, resta_macroscopic_1993, vanderbilt_electric_1993a}. Alternatively to this formulation, the spontaneous polarization is often approximated using Born effective charges that are assumed to be constant over the displacement $\Delta \mathbf{r}_i$ \cite{ghosez_dynamical_1998} 
\begin{equation}
\label{eq:spont_p_bec}
    \mathbf{P}_\text{spont} = \mathbf{P}_\text{bulk,pol} - \mathbf{P}_\text{bulk,npol} = \frac{1}{\Omega} \sum_i Z_{i}^*  \Delta \mathbf{r}_i.
\end{equation}
The Born effective charge of atom $i$ along lattice direction $j$ is defined as the derivative of the polarization with respect to the position at zero field $Z^*_{ij} = \frac{\Omega}{e}\frac{\delta P_i}{\delta r_j}|_{\varepsilon = 0}$. Note that the use of Born effective charges in Eq. \ref{eq:spont_p_bec} is valid in the bulk where the electric field is zero but not necessarily in a ferroelectric slab where the electric field usually is nonzero \cite{ghosez_dynamical_1998}.

The surface charge of the BaTiO$_3$ slab is calculated analogously to Eq.~\ref{eq:interface_theorem} as the component of the bulk spontaneous polarization along the surface normal \cite{vanderbilt_electric_1993a}. The resulting charges are illustrated in purple in the top and middle of Fig.~\ref{fig:illu_all}b. If these charges are not screened, they lead to a depolarizing field proportional to the magnitude of the polarization in the film \cite{ghosez_firstprinciples_2006}. For screening purposes and device applications, ferroelectrics are therefore often interfaced with a metallic electrode such as SrRuO$_3$.

However, SrRuO$_3$ is not an ideal metal and can only provide imperfect screening. The metal only manages to screen the interface charges after a finite distance inside the electrode, characterized by the effective screening length $\lambda_\text{eff}$. The screening length can be interpreted as the finite distance between the bound charges at the interface (purple +/– at the top of Fig.~\ref{fig:illu_all}b) and the net compensating metallic screening charges in the electrodes (white +/–) \cite{ghosez_firstprinciples_2006, stengel_band_2011}.

The finite screening length leads to a critical thickness below which the spontaneous polarization is absent in BaTiO$_3$ \cite{junquera_critical_2003, kim_critical_2005, deluca_nanoscale_2017}. The finite screening length also means that the concentration of screening carriers close to the interface decays gradually as shown in yellow in the middle of Fig.~\ref{fig:illu_all}b. The distance between the bound (purple) and screening (white) charges sets the interface dipole, which generates a voltage drop at the interface, resulting in a depolarizing field $\mathcal{E}_d$ opposite to the spontaneous polarization inside the ferroelectric \cite{ghosez_firstprinciples_2006, black_electricfield_1999, mehta_depolarization_1973} 
\begin{equation}
    \mathcal{E}_d = -\frac{2 \lambda_\text{eff}}{d \epsilon_0} P_{\text{spont},j}.
\end{equation} 
The depolarizing field depends on the size of the polarization $P_{\text{spont},j}$ along the surface normal, the thickness $d$ of the film and the effective screening capability of the metal $\lambda_\text{eff}$ and vanishes for perfect screening ($\lambda_\text{eff} \rightarrow 0$). For symmetric BaTiO$_3$/SrRuO$_3$ capacitors (Fig.~\ref{fig:symm_as_sl}a), the depolarizing field leads to a critical thickness of about 5-6 unit cells in experiments and DFT calculations \cite{junquera_critical_2003, deluca_nanoscale_2017}. 

The depolarization field also leads to a band shift in the DOS, see illustration in Fig.~\ref{fig:illu_dos}b. This band shift depends on the orientation of the spontaneous polarization, resulting in different band alignments at the two metal-ferroelectric interfaces even for symmetric capacitors. In Fig.~\ref{fig:illu_dos}b (left side, $\textbf{P}_\text{spont}$ points up), the interface dipole pointing away from the lower interface (negative bound charge) pushes the BaTiO$_3$ valence band closer to the Fermi level. For the opposite orientation of the BaTiO$_3$ polarization (right side, $\textbf{P}_\text{spont}$ points down), the interface dipole points towards the interface (positively charged), so that the conduction band is closer to the Fermi level. For BaTiO$_3$/SrRuO$_3$, the band alignment is estimated to change between \SI{0.3}{V} and \SI{1.8}{V} upon polarization reversal depending on whether relaxations of the electrode (which effectively alter the total effective screening length of the metal) are included or not \cite{stengel_band_2011, velev_magnetic_2009}.  

It is important to note that the screening length is a property of the entire interface and not only the metal. The effective screening length takes into account any contribution that reduces or improves screening beyond incomplete electronic screening. Interface-specific effects such as interfacial bonding \cite{stengel_enhancement_2009, liu_interface_2012} or polar distortions in the metal \cite{gerra_ionic_2006, puggioni_polar_2018} influence the screening length.   

First, chemical bonding can greatly influence the effective screening length of the same metal when bonded to different ferroelectrics or at different interfaces of the same ferroelectric capacitor. If Pt serves as an electrode for BaTiO$_3$, the effective screening length is almost zero due to weak interfacial bonding, while Pt with PbTiO$_3$ as the ferroelectric or BaTiO$_3$ with SrRuO$_3$ as an electrode both have much larger screening lengths \cite{stengel_band_2011}. Due to the change in chemical environment, the screening length may change for two different interfaces within the same capacitor or for different orientations of the spontaneous polarization. 

\begin{figure}
    \centering
    \includegraphics[width=\linewidth]{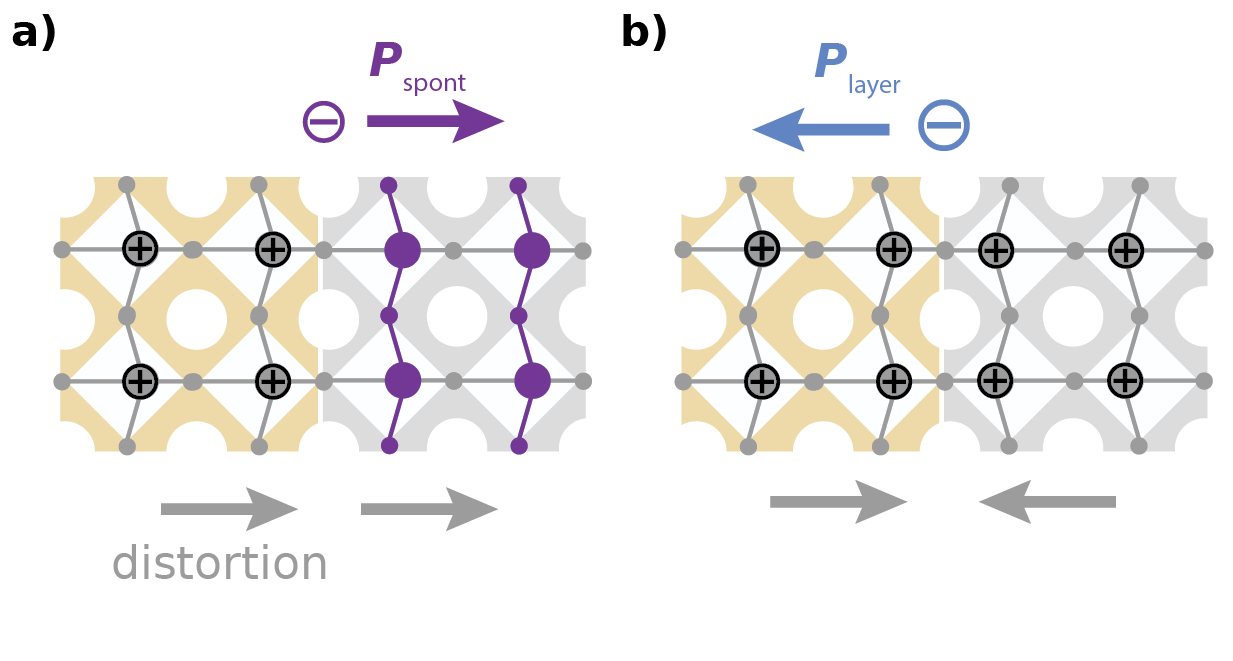}
    \caption{Induced polar distortions (gray arrows) at the metal-ferroelectric interface due to a negative bound charge at the interface from a) the ferroelectric polarization (purple) or b) formally charged layers in the metal (blue). Only the distortions of the B atom (black plus sign) are shown in the metal (yellow) or ferroelectric (gray).}
    \label{fig:pol_dists}
\end{figure}

Second, polar distortions in the metal can further reduce the effective screening length \cite{gerra_ionic_2006, puggioni_polar_2018, chisholm_atomicscale_2010, sai_ferroelectricity_2005}. Polar distortions over 2-3 formula units inside the SrRuO$_3$ electrode reduce the effective screening length when interfaced with BaTiO$_3$ \cite{gerra_ionic_2006}. Ref. \cite{puggioni_polar_2018} suggested that polar metals, i.e. metals that have polar distortions, as electrodes for ferroelectrics could even lead to zero critical thickness for the ferroelectric polarization. Such a vanishing critical thickness for the BaTiO$_3$ polarization was found in DFT calculations if the octahedral rotations of SrRuO$_3$ were suppressed and polar distortion were induced in the electrodes \cite{puggioni_polar_2018}. These distortions are usually a continuation of the ferroelectric distortion inside the metallic electrode and therefore in-phase with the ferroelectric polarization \cite{chisholm_atomicscale_2010}, see Fig. \ref{fig:pol_dists}.  

Lastly, stoichiometric BaTiO$_3$ capacitors or films with a vacuum interface have an additional asymmetry from the interfaces (Fig.~\ref{fig:symm_as_sl}b) that can further influence properties such as the critical thickness. Since the two interfaces have different chemical compositions, they have different bonding and band alignment \cite{velev_magnetic_2009}. For the case of BaTiO$_3$/SrRuO$_3$, the chemical bonding at the RuO$_2$-BaO interfaces pins the BaTiO$_3$ polar displacements in one direction, further enhancing the asymmetry and favoring one polarization direction over the other \cite{liu_interface_2012, gerra_ferroelectricity_2007}. The preferred polarization direction has a smaller critical thickness and a larger polarization value for thin films compared to the opposite direction \cite{liu_interface_2012, gerra_ferroelectricity_2007}. Still, due to imperfect screening of the metallic electrode, the polarization is suppressed for thin films even in the preferred direction and obtains its full bulk polarization value only for thick slabs \cite{liu_interface_2012}. While observable, these effects are usually much smaller than the asymmetry introduced from interface charges originating from the layer or spontaneous polarization (see appendix for more details).

In this section, we examined how metallic electrodes can screen interface charges that arise from spontaneous polarization. We reviewed the description of incomplete screening by an effective screening length and addressed electronic and chemical contributions to the screening process. Even though the interface between BaTiO$_3$ and metallic SrRuO$_3$ is formally charge neutral, the effect of the interface dipoles on both materials strongly resembles that of a polar discontinuity between two insulators. 

\subsection{Interface between BiFeO$_3$ and SrRuO$_3$}
In this section, we focus on BiFeO$_3$ where the previously discussed spontaneous and layer polarizations co-exist and interact with each other \cite{begon-lours_factors_2018, efe_happiness_2021, spaldin_layer_2021}. BiFeO$_3$ has been widely studied for its multiferroic properties \cite{wang_epitaxial_2003}, combining antiferromagnetism with a spontaneous polarization of \SI{\sim 90}{\micro\coulomb\per\cm\squared} oriented along [111] \cite{neaton_firstprinciples_2005}. BiFeO$_3$ is also a III-III perovskite with negatively charged (FeO$_2$)$^-$ and positively charged (BiO)$^+$ (001) sublayers. Again, II-IV  SrRuO$_3$ with its charge-neutral (001) layers is our choice of electrode and we will focus on stoichiometric BiFeO$_3$/SrRuO$_3$ superlattices with asymmetric interfaces. 

The total bulk polarization of BiFeO$_3$, as defined in Eq.~\ref{eq:P_layer}, consists of contributions from the spontaneous polarization and the layer polarization
    \begin{equation}
        \mathbf{P}_\text{bulk} = \mathbf{P}_\text{layer} + \mathbf{P}_\text{spont}.
    \end{equation} 
The projection of the spontaneous polarization onto the [001] axis leads to a polarization of \SI{\sim \pm 50}{\micro\coulomb\per\cm\squared} \cite{spaldin_layer_2021, efe_happiness_2021}, while the polar discontinuity between centrosymmetric BiFeO$_3$ and SrRuO$_3$ coincidentally has the same value of \SI{\sim + 50}{\micro\coulomb\per\cm\squared} and is oriented from the FeO$_2$-SrO to the BiO-RuO$_2$ interface.

If BiFeO$_3$ has an unreconstructed cleaved interface to vacuum or another layer-neutral material, its two polarization directions are no longer equivalent due to the nonzero layer polarization  \cite{spaldin_layer_2021, efe_happiness_2021}. If the spontaneous polarization is opposite to the layer polarization (Fig.~\ref{fig:illu_all}c), the spontaneous polarization provides the required \SI{0.5}{\elementarycharge\per \textit{S}} for screening and the interface charges are formally compensated. This orientation, referred to as the `happy' orientation in the literature, is energetically favored \cite{efe_happiness_2021, spaldin_layer_2021}. The spontaneous and the layer polarization add up to an interface charge of \SI{\sim + 100}{\micro\coulomb\per\cm\squared} in the opposite, `unhappy' orientation (Fig.~\ref{fig:illu_all}d) which is energetically highly disfavored \cite{efe_happiness_2021, spaldin_layer_2021}.

The SrRuO$_3$ electrode only needs to screen uncompensated interface charges in the unhappy orientation of the BiFeO$_3$ spontaneous polarization. Analogous to BaTiO$_3$/SrRuO$_3$, the screening by SrRuO$_3$ is incomplete, resulting in an interface dipole in SrRuO$_3$ at the interface and a depolarizing field in the BiFeO$_3$ slab, see Fig.~\ref{fig:illu_all}d. While most concepts discussed for BaTiO$_3$/SrRuO$_3$ also apply for BiFeO$_3$/SrRuO$_3$, there are some differences in the behavior of these two systems: First, the preference of the spontaneous polarization of BiFeO$_3$ for the happy orientation is substantially more pronounced, with calculated energy differences between the two polarization orientations being one to two orders of magnitude higher for BiFeO$_3$/SrRuO$_3$ \cite{spaldin_layer_2021} compared to asymmetric BaTiO$_3$/SrRuO$_3$ \cite{liu_interface_2012, velev_magnetic_2009}, see Tab.~\ref{tab:delta_e_pol} in the appendix. Second, BiFeO$_3$ displays a reduced polarization primarily in the unhappy orientation, while maintaining the bulk polarization value in the happy orientation also for thin slabs \cite{spaldin_layer_2021, yu_interface_2012}.

The potential difference between the two interfaces again strongly depends on the orientation of the spontaneous polarization \cite{spaldin_layer_2021}, see Fig.~\ref{fig:illu_dos}c. When BiFeO$_3$ is paraelectric, the potential difference between the two BiFeO$_3$/SrRuO$_3$ interfaces due to the discontinuity between the layer polarizations alone was estimated to be \SI{\sim 1.5}{V} \cite{yu_interface_2012}. The asymmetry coming from the differently charged interfaces in BiFeO$_3$/SrRuO$_3$ is larger than that originating from the different interfacial chemistries in asymmetric BaTiO$_3$/SrRuO$_3$ discussed earlier. The spontaneous polarization further increases (unhappy orientation) or reduces (happy orientation) this potential difference. In the happy orientation, the electric field in BiFeO$_3$ is minimal, the potential is constant and the Fermi level is approximately in the middle of the BiFeO$_3$ band gap throughout the entire slab, see left side of Fig.~\ref{fig:illu_dos}c. In the unhappy direction (right side), there is a large electric field in the slab and a large potential difference between the interfaces \cite{spaldin_layer_2021}. As a result, the Fermi level is close to/in the conduction band at the positive (BiO)$^+$-RuO$_2$ interface and is close to/in the valence band at the negative (FeO$_2)^-$-RuO interface \cite{stengel_band_2011}. This is in contrast to asymmetric BaTiO$_3$/SrRuO$_3$, where the band offset between the interfaces has a similar size in both polarization directions \cite{velev_magnetic_2009}. 

In this section, we have reviewed how the interplay of spontaneous and layer polarization in BiFeO$_3$ leads to the favoring of one polarization direction. In addition to BiFeO$_3$ other III-III and I-V ferroelectric perovskites such as KNbO$_3$ or LiNbO$_3$ display a similar interplay of spontaneous and layer polarization. An extensive list of experimentally accessible materials with formally charged layers is provided in Ref. \cite{efe_engineering_2024}.

\section{Charged layers in the metallic slab: The example of \texorpdfstring{\NoCaseChange{SmNiO$_3$}}{SmNiO3}}

Surprisingly, results consistent with a polar discontinuity have also been observed when the formally charged layers are within the metallic slab, such as in SmNiO$_3$ \cite{simmen_interplay_2025, efe_polarizing_2026}. 

SmNiO$_3$ is a member of the rare-earth nickelate family which have a low-temperature insulating $P2_1/c$ (except for LaNiO$_3$) and a high-temperature metallic $Pbnm$ phase \cite{rodriguez-carvajal_neutrondiffraction_1998, lacorre_synthesis_1991}. In the metallic phase of SmNiO$_3$, a formal charge of +3 is usually assigned to Sm and Ni \cite{lacorre_synthesis_1991, catalano_rareearth_2018}. Assuming these formal charges, SmNiO$_3$ has (001) layer charges of \SI{\pm 1}{\elementarycharge} and a polar discontinuity of \SI{50}{\micro\coulomb\per\cm\squared} at the interface to paraelectric BaTiO$_3$. This polar discontinuity would be exactly screened by the spontaneous polarization of BaTiO$_3$ of \SI{\pm50}{\micro\coulomb\per\cm\squared} when strained to SrTiO$_3$ lattice parameters \cite{simmen_interplay_2025}. But since SmNiO$_3$ is metallic, neither the layer charges nor the polar discontinuity are formally defined, which raises the question of how these observations can be rationalized.

In this section, we examine this behavior using the example of SmNiO$_3$/BaTiO$_3$.
We first give an overview of signatures of such a formal-charge polar discontinuity in DFT calculations of heterostructures with either the NiO$_2$-BaO or SmO-TiO$_2$ interface, before rationalizing the results using the concept of an interface dipole.

\subsection{Signatures of a formal-charge polar discontinuity in SmNiO$_3$/BaTiO$_3$}

\begin{figure*}
    \centering
    \includegraphics[width=\textwidth]{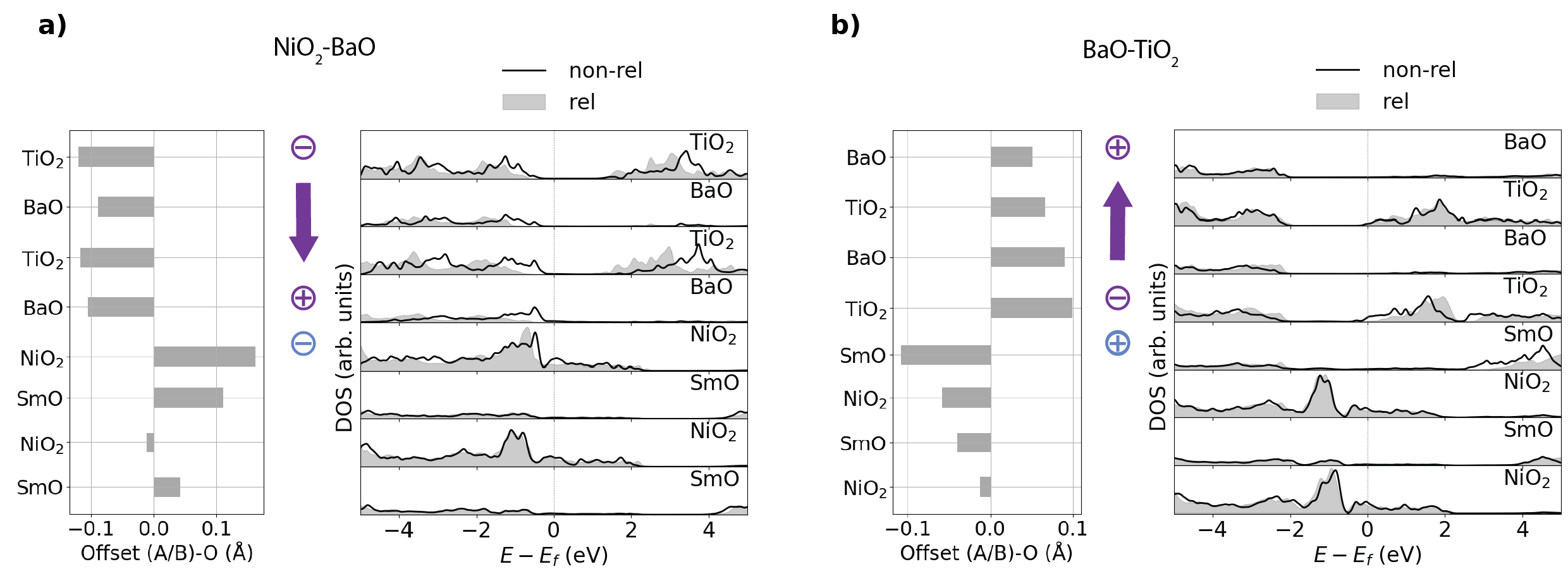}
    \caption{Polar displacements and the layer-projected DOS in the sublayers around the NiO$_2$-BaO (a) and SmO-TiO$_2$ (b) interfaces in freestanding (BaTiO$_3$)$_4$(SmNiO$_3$)$_6$. The formal interface charge is indicated in blue and the direction of the spontaneous polarization in the relaxed heterostructures in purple. }
    \label{fig:data_bto_sno}
\end{figure*}

Experimental and computational studies of metallic SmNiO$_3$ interfaced with ferroelectric BaTiO$_3$ strongly suggest the presence of a formal-charge polar discontinuity in BaTiO$_3$/SmNiO$_3$ \cite{simmen_interplay_2025, efe_polarizing_2026}. First, the spontaneous polarization of BaTiO$_3$ shows a strong preference for the polarization orientation that would minimize the formal-charge polar discontinuity \cite{simmen_interplay_2025, efe_polarizing_2026}. During film growth, the direction of the BaTiO$_3$ spontaneous polarization was found to be consistent with such an interfacial bound charge  \cite{efe_polarizing_2026}, and in DFT relaxations initiated from the paraelectric phase, the spontaneous polarization always adopts the `happy' orientation \cite{simmen_interplay_2025}. Fig.~\ref{fig:data_bto_sno} (left figure in a and b) shows the polar distortion of BaTiO$_3$ for the NiO$_2$/BaO (a) and SmO-TiO$_2$ (b) interface in relaxed (BaTiO)$_4$/(SmNiO$_3$)$_6$ heterostructures after initiation to the paraelectric structure in DFT. The difference in Ba-O positions is negative at the NiO$_2$-BaO interface, with the Ba lying closer to the interface compared to the O and the spontaneous polarization pointing towards the interface, minimizing the formal-charge polar discontinuity. Conversely, at the BaO-NiO$_2$ interface, the oxygen atoms are closer to the interface than the Ti (positive offset) and the spontaneous polarization points away from the interface, again adopting the `happy' orientation. 

The spontaneous polarization of BaTiO$_3$ in the happy orientation was found to be stable down to one unit cell in DFT calculations and experiments, with polarization values very close to the bulk, indicating the absence of a critical thickness in the `happy orientation \cite{simmen_interplay_2025, efe_polarizing_2026}. Analogous to BiFeO$_3$/SrRuO$_3$, the opposite orientation is predicted to be energetically strongly disfavored (`unhappy') and metastable only for thicker BaTiO$_3$ slabs \cite{simmen_interplay_2025}.

Additional polar displacements opposite to those of the spontaneous polarization are present in the SmNiO$_3$ electrode if rotations are suppressed at the interface (see Fig.~\ref{fig:data_bto_sno}c and d) or throughout the entire slab \cite{simmen_interplay_2025}. Although polar displacements commonly exist in metal-oxide electrodes, they are usually a continuation of the polar mode into the electrode and in-phase with the ferroelectric polarization \cite{puggioni_polar_2018, chisholm_atomicscale_2010, liu_interface_2012, gerra_ionic_2006}, see Fig. \ref{fig:pol_dists}a. The pattern of displacements in SmNiO$_3$ is opposite to that of the spontaneous polarization in BaTiO$_3$, and is consistent with the formation of a dipole where the Ni$^{3+}$ and Sm$^{3+}$ cations are attracted to the negative bound charge at the NiO$_2$-BaO (see Fig. \ref{fig:pol_dists}b), and repelled by the positive charge at the SmO-TiO$_2$ interface. Analogous distortions have also been observed experimentally in LaNiO$_3$ at the LaNiO$_3$/PZT interface \cite{hwang_polarizationcontrolled_2025}.

Lastly, the band alignment for the two interfaces in Fig.~\ref{fig:data_bto_sno} is very different when BaTiO$_3$ is paraelectric, see black line in Fig.~\ref{fig:data_bto_sno}a  for the NiO$_2$-BaO interface and \ref{fig:data_bto_sno}b for the SmO-TiO$_2$ interface (or Fig. 3 in Ref.  \cite{simmen_interplay_2025}). Differences in band alignment of this magnitude usually occur upon polarization switching \cite{stengel_band_2011} or between the differently charged interfaces in LaAlO$_3$/SrTiO$_3$ \cite{bristowe_oxide_2009}, and are much larger than expected from just a chemical asymmetry as in stoichiometric BaTiO$_3$/SrRuO$_3$ capacitors (see appendix for more details). 
 
\subsection{The origin of the apparent formal-charge polar discontinuity}

Despite results consistent with formally charged layers and a polar discontinuity, we cannot directly apply the principles discussed for BiFeO$_3$/SrRuO$_3$ and LaAlO$_3$/ SrTiO$_3$ here. The polarization is ill-defined in the presence of free metallic carriers and the modern theory of polarization does not formally apply. We therefore rationalize the observations made in the previous section, by starting in the limit of insulating SmNiO$_3$, and discussing what formally changes when metallic carriers are introduced. We first explain how imperfect screening of an underlying charged lattice in the metallic slab leads to the observed signatures of an apparent formal-charge polar discontinuity, before discussing factors determining the size of the interface dipole resulting from this imperfect screening. 

\subsubsection{Large interface dipole in the formally charged, metallic SmNiO$_3$ slab} 

\begin{figure}
    \centering
    \includegraphics[width=\linewidth]{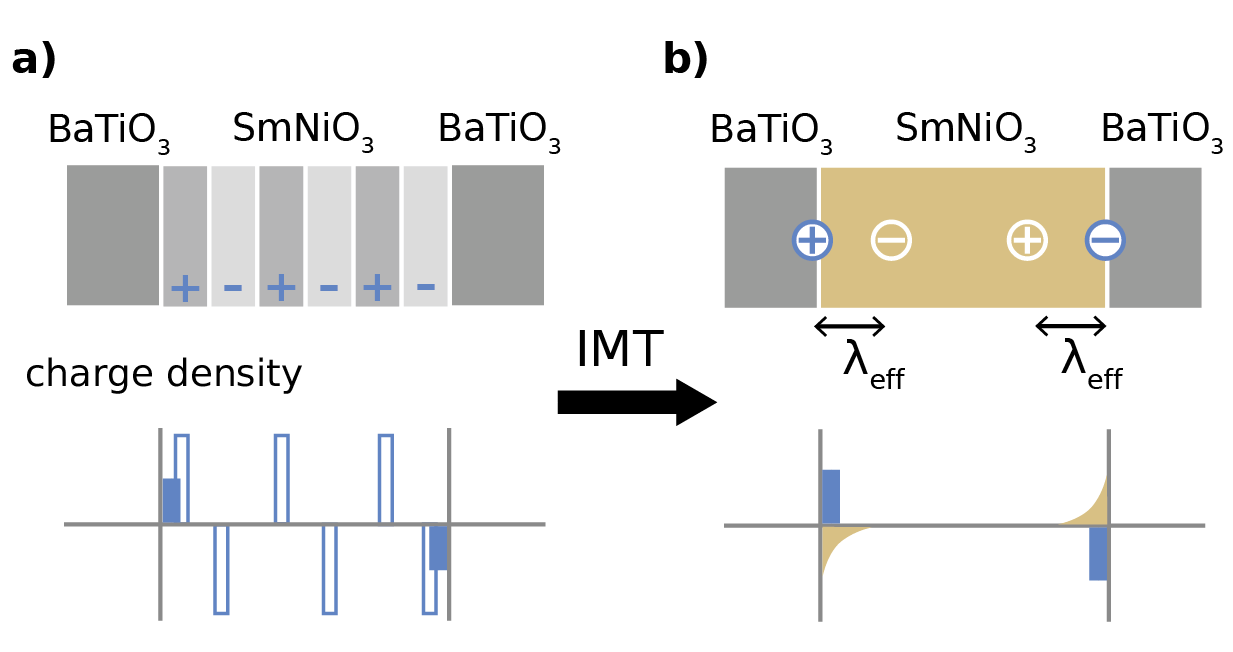}
    \caption{Illustration of the formal-charge polar discontinuity at the interface between paraelectric BaTiO$_3$ and SmNiO$_3$ in the a) insulating and b) metallic phase. Metallic carriers and metallic SmNiO$_3$ are shown in yellow, insulating BaTiO$_3$ in gray and bound charges from the formal-charge layer polarization in blue.}  
    \label{fig:illu_bto_sno}
\end{figure}

In the insulating phase, SmNiO$_3$ has (001) layers with a charge of \SI{\pm 1}{\elementarycharge\per\textit{S}} analogous to LaAlO$_3$, see the blue +/-- in Fig.~\ref{fig:illu_bto_sno}a and the corresponding charge density below (blue empty bars). Since the alternatingly charged layers compensate each other within the SmNiO$_3$ slab, uncompensated charges are only present at the interface to paraelectric BaTiO$_3$. To study the effect of the resulting polar discontinuity on the system, we can thus equivalently consider only the bound charge (blue filled bar) at the interface instead of each individual layer charge (blue empty bars) within the SmNiO$_3$ slab. For the interface of insulating SmNiO$_3$ to paraelectric BaTiO$_3$, this bound charge of \SI{\pm 0.5 }{\elementarycharge\per\textit{S}} corresponds to the total interface charge. 

Following the insulator-to-metal transition (IMT), free screening charges accumulate close to the interface in the SmNiO$_3$ (yellow), and the resulting picture strongly resembles that of the BaTiO$_3$/SrRuO$_3$ interface (see Fig. \ref{fig:illu_all}b), except that the screening and bound charges are in the same material. Due to the finite screening length of the metal, the screening charges (white +/-- in Fig.~\ref{fig:illu_bto_sno}b) are a finite distance from the bound charge at the interface (blue +/-- in Fig.~\ref{fig:illu_bto_sno}b), creating an interface dipole (dipole created by blue and white +/--) \cite{burton_evolution_2010, lione_first_2025}. Since there are usually enough screening charges inside the metallic slab to compensate for the bound interface charges, the two interfaces are formally charge neutral. (We will continue to refer to the `bound charge' or the `underlying charged lattice' in the metallic phase even though, formally, any charge in the metal would be screened by the free carriers.)

The experimental and computational results can be explained by this interface dipole: The interface dipole points in the same direction as the layer polarization in the insulating SmNiO$_3$ phase, polarizing the adjacent layer. The interface dipole shifts the bands in opposite directions at the two interfaces, resulting in a large difference in band alignment. The BaTiO$_3$ prefers the direction of the spontaneous polarization that aligns with the interface dipole, while the induced polar displacements in the SmNiO$_3$ result from an attraction/repulsion of the ions towards/away from the bound charge and/or the screening charges.

While interface dipoles are a general property of an interface with metallic screening charges, we emphasize that the interface dipoles in systems like BaTiO$_3$/SmNiO$_3$ are unusually large. To highlight this, we provide a direct comparison between BaTiO$_3$ with SmNiO$_3$ or SrRuO$_3$ electrodes in the appendix.     

In general, the magnitude of the interface dipole depends on the size of the bound charge at the interface and the effective screening length of the metal. For a good metal or an interface with a small bound charge, we expect a smaller interface dipole and a smaller influence on the overall system than for a bad metal or an interface with a large bound charge. If the magnitude of the interface dipole per unit area is similar to the size of the polar discontinuity between insulating SmNiO$_3$ and paraelectric BaTiO$_3$, the metal might appear to not be screening the polar discontinuity at all. In the next sections, we will discuss the contributions of the bound charge and the screening length to the interface dipole in more detail.  

Since the interface in the interface-dipole picture is formally charge neutral, there is no polar discontinuity at the interface between SmNiO$_3$/BaTiO$_3$. Still, it can be intuitive to think of this interface in such a polar-discontinuity picture. In this hypothetical picture, the screening charges would merely reduce the interface charge of \SI{\pm 0.5 }{\elementarycharge\per\textit{S}} resulting from the formally charged layers. The observations would then be explained analogously to BiFeO$_3$: The direction of the BaTiO$_3$ polarization minimizing the interface charges are energetically preferred (`happy' orientation), and the opposite interface charges lead to a large change in band alignment between the two possible BaTiO$_3$/SmNiO$_3$ interfaces. 

\subsubsection{How does the interfacial bound charge differ between insulating and metallic SmNiO$_3$?}

\begin{figure*}
    \centering
    \includegraphics[width=\linewidth]{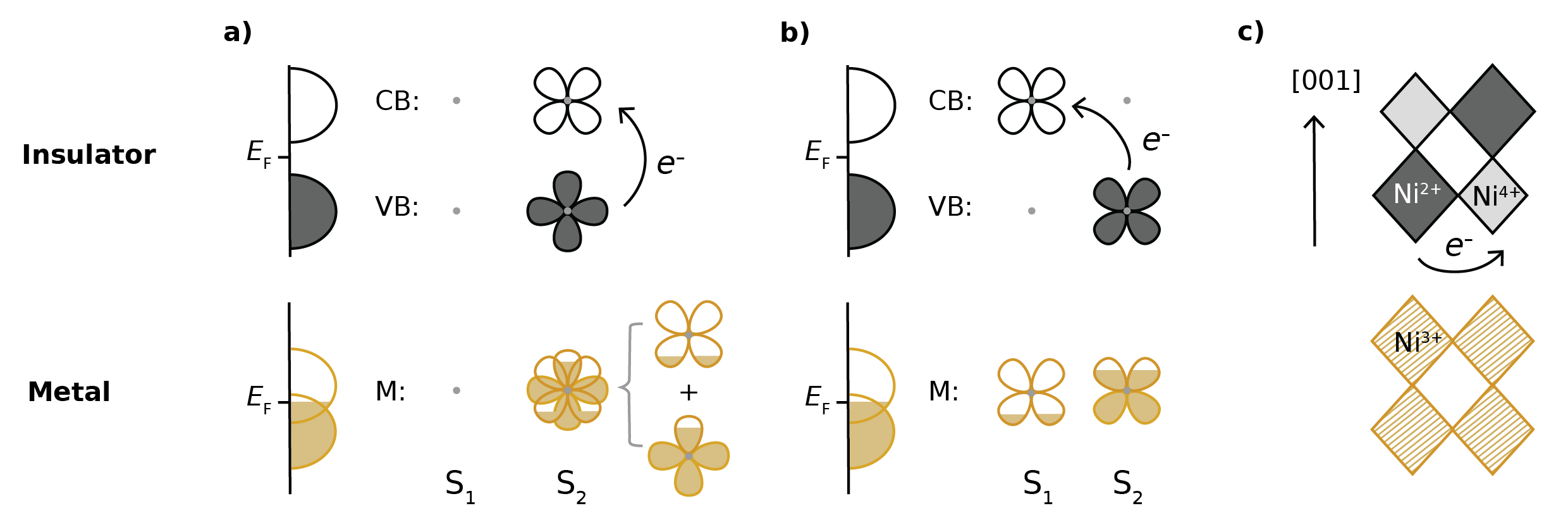}
    \caption{Illustration of the Wannier function occupation between the insulator (gray) to metal (yellow) phases where the conduction bands (CB) and valence bands (VB) consist of states on a) the same site (S$_2$) or b) on different sites (S$_1$ and S$_2$, respectively). c) The formal layer charge does not change across the IMT in SmNiO$_3$ due to the checkerboard arrangement of the two distinct Ni sites.}
    \label{fig:illu_mit_bound_charge}
\end{figure*}

In this section, we qualitatively discuss how the bound charge differs between the insulating and metallic SmNiO$_3$ phases based on the occupation of Wannier functions. Based on our argument, we expect the bound charge to change its value across an IMT only if the gap is closed by bands located on different crystal sites. For the sake of simplicity, we use the same Wannier functions with different occupations for the insulating and metallic phases. Furthermore, we assume that it is justified to draw a link between our argument and the magnitude of the bound charge at the interface, as the layer charge of an insulator is rigorously defined using Wannier functions (see Eq. \ref{eq:P_layer}). 

For systems where the gap is closed by combining bands on the same site (Fig. \ref{fig:illu_mit_bound_charge}a), such as in Mott insulators or systems with a Jahn-Teller distortion, the bound charge is likely unaffected by the IMT. In the insulating phase, we can describe the valence and conduction bands by a fully occupied (filled gray) and completely empty (empty black) Wannier function located on site $S_2$. In the metallic phase, these bands overlap and electrons (filled yellow area) are transferred from one to the other Wannier function. Since both Wannier functions are centered around the same site ($S_2$), we do not expect the formal charge of the atom and consequently the formal layer charge and the magnitude of the bound charge to change across the IMT.

In charge-transfer insulators or a gap opened by charge disproportionation, the nature of the valence and conduction bands is different. In the simplest picture, the filled Wannier function (gray) corresponding to the valence band is located on site $S_2$ (Fig. \ref{fig:illu_mit_bound_charge}) while the conduction band corresponds to a completely empty Wannier function on site $S_1$. In the metallic phase, these bands again overlap around the Fermi level and electrons (filled area) are partially transferred from site $S_2$ to $S_1$. This results in a reduction of the formal charge of site $S_1$ by the amount of electrons transferred and an increase of the formal charge on site $S_2$ by the same amount (white area). Depending on the arrangement of these crystal sites with respect to the interface, this type of IMT could thus change the underlying charged lattice of the metal and the magnitude of the bound charge.

In reality, the bands do not consist of a single atom or orbital type and the bands around the Fermi level in SmNiO$_3$ are formed by strongly hybridized Ni-$d$ and O-$p$ states \cite{varignon_complete_2017}. 
Here, the relevant electron transfer during the IMT occurs between two neighboring NiO octahedra, see Fig. \ref{fig:illu_mit_bound_charge}c. The gap is closed as the breathing distortion on the Ni sites disappears and the electrons are distributed equally between the Ni sites. Consequently, the charges centered around these two Ni sites are likely different in the metallic and insulating phases. However, since the electrons are transferred within one (001) layer (arrow in Fig. \ref{fig:illu_mit_bound_charge}c), we expect the formal layer charge and the bound charge at the interface to be unaffected by the IMT. Note that, for a different choice of interface plane, the formal layer charge likely still changes across the IMT.

The existing literature further supports our qualitative arguments. In DFT+DMFT calculations of rare-earth nickelates, the transition from the metallic to the insulating state is commonly described by maximally localized Wannier functions (MLWF) that remain centered around the Ni atoms throughout the transition \cite{subedi_lowenergy_2015, hampel_energetics_2019} and which resemble each other in the metallic and insulating phases \cite{subedi_lowenergy_2015}. In these calculations, the $e_\text{g}$ bands in the metallic phase are quarter-filled, consistent with the Ni$^{3+}$ or $d^8L$ configuration \cite{subedi_lowenergy_2015}. Furthermore, an analysis of the amount of screening charges in a similar system, NdNiO$_3$/SrTiO$_3$, showed a difference between the NdO-TiO$_2$ and NiO$_2$-SrO interfaces consistent with a bound charge of \SI{\pm 0.5 }{\elementarycharge\per\textit{S}}  \cite{lione_first_2025}.

Note that SmNiO$_3$ together with the other rare-earth nickelates is usually referred to as a negative-charge transfer insulator, where the Ni ions are better described as being $d^8$ with a ligand-hole on oxygen instead of having a +3 formal charge \cite{park_siteselective_2012, bisogni_groundstate_2016}. This negative charge transfer is, however, not relevant for the argument presented in this section because it relates to the description of the covalent bonding between Ni and O. The ligand-hole picture is an alternative way to describe the same physics as the charge disproportionation picture \cite{carta_importance_2026}. 
As we discussed in the beginning for LaAlO$_3$/SrTiO$_3$, a change in covalency does not affect the formal layer charge in a perovskite (see Section \ref{sec:player_lao}) \cite{bristowe_net_2011}.

We conclude that, for the [001] direction, the IMT likely does not substantially affect the underlying charged lattice in SmNiO$_3$ and the bound charge of \SI{\pm 0.5 }{\elementarycharge\per\textit{S}} at the interface to paraelectric BaTiO$_3$ persists in the metallic phase. 

\subsubsection{What is the influence of the metallic screening length?}
In addition to the amount of bound and screening charges, the magnitude of the interface dipole is also determined by their spatial separation ($\lambda_\text{eff}$ in Fig. \ref{fig:illu_bto_sno}b). For better metals with a shorter screening length, the bound and screening charges are closer together, resulting in a smaller interface dipole and less distinct features of an apparent formal-charge polar discontinuity. For bad metals with a large screening length, the separation of the charges, the interface dipole, and the observed experimental and computational signatures are larger. 

Transition metal oxides such as SmNiO$_3$ are usually considered bad metals, and their resistivities are one to two orders of magnitude higher than those of typical elemental metals \cite{hickox-young_polar_2023}, and consequently, we find the signatures described earlier. However, it is still surprising how closely these results resemble those of a polar discontinuity between two insulators \cite{simmen_polar_2026}. To better understand the role of the finite screening length, we therefore propose future studies of the effect of layer charges in metals with varying screening capabilities, e.g. by influencing the correlations or the bandwidth in the material, or by studying trends across systems with different conductivities.

\section{Summary and conclusion}
In the first part of this work, we reviewed the current understanding of polar discontinuities and their screening in perovskite superlattices using the representative systems LaAlO$_3$/SrTiO$_3$, BaTiO$_3$/SrRuO$_3$ and BiFeO$_3$/SrRuO$_3$.  Starting with LaAlO$_3$/SrTiO$_3$,  we explained how a non-zero layer polarization exists in systems with a half-quantum-containing polarization lattice (LaAlO$_3$), leading to a polar discontinuity at the interface with materials with a zero-containing polarization (SrTiO$_3$) that can be screened by forming a 2DEG. Second, we discussed the (imperfect) screening of the bound charges arising from the ferroelectric polarization by the metallic electrode in BaTiO$_3$/SrRuO$_3$. Third, we summarized how the layer and spontaneous polarization of BiFeO$_3$ co-exist and influence each other in BiFeO$_3$/SrRuO$_3$, resulting in a preferential `happy' orientation of the spontaneous polarization.

Next, we presented evidence of behavior consistent with formally charged layers in metallic SmNiO$_3$, similar to those of BiFeO$_3$ or LaAlO$_3$. The strong interface dependence of the band alignment between SmNiO$_3$ and BaTiO$_3$, the existence of a preferential BaTiO$_3$ spontaneous polarization direction and polar distortions inside the metal adjacent to the interface opposite to those of the BaTiO$_3$   strongly suggest the existence of a bound charge that originates from an underlying charged lattice in metallic SmNiO$_3$. We rationalized these effects using the concepts introduced in the first part by considering the consequences of incomplete screening at the metal-ferroelectric interface. The finite screening length of the metal leads to an interface dipole that can rationalize the experimental and computational findings.  

It remains surprising how well the observations for systems with formally charged layers in the metal agree with similar studies for systems where the charged layers are located in the insulators. This indicates that the metallic electrons in SmNiO$_3$ remain largely localized at their parent ions and raises the question whether the `free' electrons in SmNiO$_3$ are better described as `quasi-bound'. In that case, a description of the interface in terms of a polar discontinuity and a connection to the modern theory of polarization is quite intuitive.

We hope this work inspires future research that addresses the intriguing questions posed here. 

\section{Acknowledgements} 

This research was funded by the Swiss National Science Foundation (Grant No. 209454) and ETH Zurich. Computational work was carried out on the Swiss National Supercomputing Center’s Daint cluster under Project No. lp61.\\

\section{Data availability}
The input files and datasets relevant to our ab initio calculations are publicly accessible via the Materials Cloud Archive at \url{http://doi.org/DOItobeinsertedhere}.

\section{APPENDIX A: Methods}

Density functional theory (DFT) calculations were carried out using the VASP package \cite{kresse_efficiency_1996, kresse_efficient_1996}, employing the PBEsol exchange-correlation functional \cite{perdew_restoring_2008}. Projector augmented-wave (PAW) pseudopotentials \cite{blochl_projector_1994, kresse_ultrasoft_1999} with the following valence electron configurations were used: 5$s^2$ 5$p^6$ 5$d^{0.001}$ 6$s^{1.999}$ (Ba), 3$p^6$ 3$d^9$ 4$s^1$ (Ni), 3$s^2$ 3$p^6$ 3$d^3$ 4$s^1$ (Ti), 2$s^2$ 2$p^4$ (O), and 5$s^2$ 5$p^6$ 5$d^1$ 6$s^2$ (Sm), with the $f$ electrons frozen in the core for Sm. The superlattices and heterostructures were constructed from an in-plane $\sqrt{2} \times \sqrt{2}$ supercell with 6 formula units of SmNiO$_3$ and 5 of BaTiO$_3$ along the out-of-plane direction. The in-plane lattice constant were strained to those of SrTiO$_3$ with pseudo-cubic lattice parameter $a = \SI{3.898}{\AA}$ (DFT value from Ref.~\cite{wahl_srtio3_2008}) to simulate experimental substrate conditions.  We used a cutoff energy of 650~eV alongside a $\Gamma$-centered $8\times8\times1$ $k$-point grid, and relaxed the internal coordinates until force components were smaller than $0.01~\text{eV\,\AA}^{-1}$. For relaxations of heterostructures, we included a 30~\AA\ vacuum region and employed a dipole correction to avoid interactions between the surfaces.  Additionally, the two vacuum-terminating SmNiO$_3$ layers were held fixed in their bulk geometry while the other SmNiO$_3$ layers and the BaTiO$_3$ slab were allowed to relax.

\subsection{APPENDIX B: Comparison of asymmetric BaTiO$_3$/SrRuO$_3$ and BaTiO$_3$/SmNiO$_3$}

\begin{figure}
    \centering
    \includegraphics[width=\linewidth]{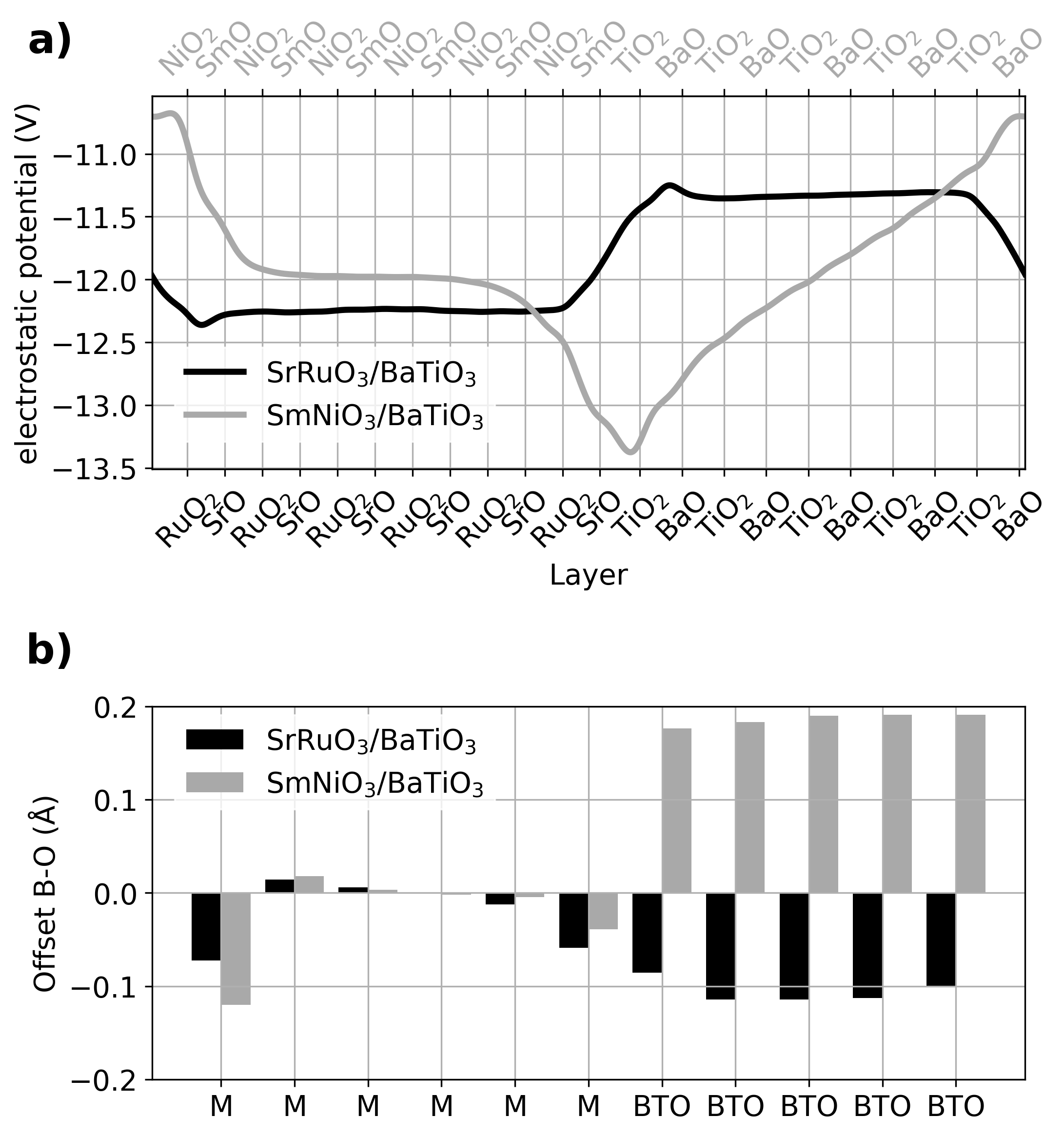}
    \caption{Comparison of asymmetric BaTiO$_3$/SrRuO$_3$ (black) and BaTiO$_3$/SmNiO$_3$ (gray). a) Macroscopic average of the potential of the electron for paraelectric BaTiO$_3$. b) Polar distortions in relaxed superlattices with the BaTiO$_3$ polarization in its preferred direction. The offset between the B-atom (B = Ti, Ru or Ni) and the oxygen plane is taken as a measure for the spontaneous polarization. }
    \label{fig:comp_bto-sro_bto-sno}
\end{figure}

Interfacial asymmetry due to a difference in chemical composition, bound charges, or both can lead to distinct features in ferroelectric capacitors, see sections on BaTiO$_3$/SrRuO$_3$, BiFeO$_3$/SrRuO$_3$ and BaTiO$_3$/SmNiO$_3$. Here, we highlight that even though these effects are present in all three systems, they are much larger if the asymmetry results from an underlying charged lattice in the ferroelectric or metal. Using examples from the literature and DFT calculations, we discuss how the band alignment and the preference for one of the two polarization orientations differ for an asymmetry originating from chemistry or bound charges.

The asymmetry in BaTiO$_3$/SmNiO$_3$ due to the interfacial bound charges is one order of magnitude larger than in stoichiometric BaTiO$_3$/SrRuO$_3$ which only has a chemical asymmetry. Fig. \ref{fig:comp_bto-sro_bto-sno}a shows the macroscopically averaged potential of the electron in BaTiO$_3$/SrRuO$_3$ (black) and BaTiO$_3$/SmNiO$_3$ (gray) superlattices with paraelectric BaTiO$_3$, with the positions of all ions constrained to those of the bulk unit cells. The asymmetry in stoichiometric BaTiO$_3$/SrRuO$_3$ with two chemically distinct interfaces leads to a difference in band alignment between the two interfaces on the order of \SI{\sim 0.1}{V} \cite{gerra_ferroelectricity_2007}. In comparison, the asymmetry due to the opposite interfacial bound charges in BaTiO$_3$/SmNiO$_3$ is one order of magnitude larger (\SI{\sim 2}{V}) \cite{simmen_interplay_2025}. As a result, we observe a steep slope in the BaTiO$_3$ potential when interfaced with SmNiO$_3$ compared to an almost flat potential of BaTiO$_3$ interfaced with SrRuO$_3$. 

The difference in band alignment between the two interfaces remains similar for both polarization orientations in BaTiO$_3$/SrRuO$_3$ but strongly depends on the direction of the polarization in BaTiO$_3$/SmNiO$_3$. In BaTiO$_3$/SrRuO$_3$, the slope of the macroscopically averaged potential in the BaTiO$_3$ slab changes only little between the two polarization orientations (See Fig. 2 in Ref. \cite{velev_magnetic_2009}). In contrast, the electric field in BaTiO$_3$ interfaced with SmNiO$_3$ differs significantly for the two polarization orientations, being close to zero in the happy orientation and on a similar scale as superlattices with paraelectric BaTiO$_3$ in the unhappy orientation (Fig. 7 in Ref. \cite{simmen_interplay_2025}). The behavior of BaTiO$_3$/SmNiO$_3$ much more resembles results for BiFeO$_3$/SrRuO$_3$ where the bound charge originates from the charged layers in BiFeO$_3$, see e.g., Fig. 9 in Ref. \cite{spaldin_layer_2021} or illustration in Fig. \ref{fig:illu_all} as a comparison. 

\begin{table*}
\centering
\begin{threeparttable}
{\renewcommand{\arraystretch}{1.2}
\begin{tabular}{l cc cc}
\toprule
Interface & \multicolumn{2}{c}{$\Delta E$ (meV/B-atom)} & \multicolumn{2}{c}{Critical thickness for $P_\text{spont}$ (f.u.)} \\
\cmidrule(lr){2-3}\cmidrule(lr){4-5}
         & Lit. & Calc. & Favored & Disfavored \\ \midrule
        BaTiO$_3$/SrRuO$_3$ \cite{liu_interface_2012, velev_magnetic_2009} & 2--7 & 9 & $\leq 4$\tnote{a} & 6--8 \\
        BaTiO$_3$/SmNiO$_3$ \cite{simmen_interplay_2025} & -- & 100 & 0 & 10 \\
        BiFeO$_3$/SrRuO$_3$ \cite{spaldin_layer_2021} & $250$ & -- & $<$ 6\tnote{a} & 6 \\ 
\bottomrule
\end{tabular}
}
\begin{tablenotes}\footnotesize
\item[a] Minimum size reported.
\end{tablenotes}
\end{threeparttable}
    \caption{Comparison of the energy difference between and the critical thickness of the two BaTiO$_3$ polarization directions for different superlattices. Note that the literature values vary in superlattice size, octahedral rotations and functionals used. The values were reported for superlattice sizes the following superlattice sizes $n_\text{FE}$/$n_\text{M}$ (and functionals): 8/4 (LDA) \cite{liu_interface_2012}, 6/5 (PBE) \cite{velev_magnetic_2009}, 6/4 (PBEsol) \cite{spaldin_layer_2021}.   }
    \label{tab:delta_e_pol}
\end{table*}

In agreement with the more symmetric change in band alignment for the two polarization directions in BaTiO$_3$/SrRuO$_3$ compared to the BaTiO$_3$/SmNiO$_3$, the energy difference between the two polarization orientations is also smaller in BaTiO$_3$/SrRuO$_3$. Tab. \ref{tab:delta_e_pol} shows a comparison of these energy differences in the literature and for the additional calculations presented here. Again, the energy differences of BaTiO$_3$/SrRuO$_3$ are approximately one order of magnitude smaller compared to similar calculations of BaTiO$_3$/SmNiO$_3$ or BiFeO$_3$/SrRuO$_3$ (note that the values from the literature should be compared with care as they were reported for different superlattice sizes and functionals). 

The energy difference between the two polarization orientations is accompanied by a difference in critical thickness, see Tab. \ref{tab:delta_e_pol}. While the critical thickness of the energetically favored polarization direction appears similar for all three systems at first glance, the evolution of the spontaneous polarization quantitatively differs with film thickness. For BaTiO$_3$ with SrRuO$_3$ electrodes, the polarization continuously decreases with film thickness for both polarization directions \cite{liu_interface_2012}. For BaTiO$_3$/SmNiO$_3$, the polarization in the happy orientation remains constant and very close to the bulk value for all reported film thicknesses, while the expected polarization only reduces with film thickness for the unhappy orientation \cite{simmen_interplay_2025}. Similarly, the magnitude of the polarization in a BiFeO$_3$ slab of \SI{6}{\text{f.u.}} thickness is approximately half the size in the unhappy compared to the happy orientation, where values close to the bulk polarization were reported throughout the entire slab \cite{spaldin_layer_2021}. 

The interfacial bound charges from the underlying charged lattice lead to a polarizing field for the happy polarization orientation in BaTiO$_3$/SmNiO$_3$. The bound charge from the spontaneous polarization leads to a depolarizing field for the BaTiO$_3$ polarization in both directions with a SrRuO$_3$ electrode and only in the unhappy direction for BaTiO$_3$ with a SmNiO$_3$ electrode. As a result, the spontaneous polarization in the preferred direction is expected to be smaller for BaTiO$_3$/SrRuO$_3$ compared to BaTiO$_3$/SmNiO$_3$. Fig. \ref{fig:comp_bto-sro_bto-sno}b shows a direct comparison of the spontaneous polarization in the preferred polarization direction for relaxed superlattices with \SI{5}{\text{f.u.}} of BaTiO$_3$ interfaced with SrRuO$_3$ or SmNiO$_3$. Indeed, the displacements of the Ti with respect to the O atoms are almost double in size in SmNiO$_3$/BaTiO$_3$ (gray) compared to BaTiO$_3$/SrRuO$_3$ (black). 

In conclusion, the chemical and charge asymmetry leads to similar features in all superlattices. However, the magnitude of these features strongly depends on the presence of a bound charge at the interface.

%%%% END %%%%%%
\bibliography{references}

\end{document}